\documentclass[symmetry,article,accept,moreauthors,pdftex]{Definitions/mdpi}
\firstpage{1}
\pubvolume{13}
\issuenum{08}
\articlenumber{1432}
\pubyear{2021}
\copyrightyear{2021}
\datereceived{18 May 2021}
\dateaccepted{12 July 2021}
\datepublished{5 August 2021}
\hreflink{https://doi.org/10.3390/sym13081432} % If needed use \linebreak
\pdfoutput=1

\soulregister\citep7
\soulregister\citet7
\soulregister\ref7
\usepackage{gensymb}

\Title{On a possible origin of the gamma-ray excess around the Galactic Center}

\TitleCitation{On a possible origin of the gamma-ray excess around the Galactic Center}

\Author{Dmitry O. Chernyshov$^{1}$, Andrei E. Egorov $^1$, Vladimir A. Dogiel$^{1}$*, Alexei V. Ivlev$^2$}

\AuthorCitation{Chernyshov, D.; Egorov, A.; Dogiel, V.  Ivlev, A.}
\address{%
$^{1}$ \quad Lebedev Physical Institute, Leninskii prospect - 53, Moscow, Russia\\
$^{2}$ \quad Max-Planck-Institut fu\"r extraterrestrische Physik, 85748 Garching, Germany
}
\corres{Correspondence: dogiel@lpi.ru}

\abstract{Recent observations of gamma-rays with the Fermi Large Area Telescope (LAT) in the direction of the inner Galaxy revealed a mysterious GeV excess. Its intensity is significantly above predictions of the standard model of cosmic rays (CRs) generation and propagation with a peak in the spectrum around a few GeV. Popular interpretations of this excess are due to either spherically distributed annihilating dark matter (DM) or abnormal population of millisecond pulsars. We suggested an alternative explanation of the excess through the CR interactions with molecular clouds in the Galactic Center (GC) region. We assumed that the excess could be imitated by the emission of molecular clouds with depleted density of CRs with energies below $\sim$ 10 GeV inside. A novelty of our work is in detailed elaboration of the depletion mechanism of CRs with the mentioned energies through the ``barrier'' near the cloud edge formed by the self-excited MHD turbulence. Such depletion of CRs inside the clouds may be a reason of  deficit of gamma rays from the Central Molecular Zone (CMZ) at energies below few GeV. This in turn changes the ratio between various emission components at those energies, and may potentially absorb the GeV excess by simple renormalization of key components.}

\keyword{dark matter; gamma-ray excess; galactic center} %(List three to ten pertinent keywords specific to the article; yet reasonably common within the subject discipline.)

\begin{document}
%%%%%%%%%%%%%%%%%%%%%%%%%%%%%%%%%%%%%%%%%%
%\setcounter{section}{-1} %% Remove this when starting to work on the template.

\section{Introduction and motivation}

The physical nature of dark matter (DM) remains to be one of the biggest puzzles in a modern physics despite extensive searches of DM at colliders, direct detectors and indirect observations. The latter comprises the searches of various CR particles \citep[see e.g.][]{adr17} and emissions from DM annihilation or decay, particularly in gamma rays \citep[e.g.][part \MakeUppercase{\romannumeral 5}]{Bertone-book}. Possible gamma-ray signals from DM may have in turn two different spectral types: wideband continuum and narrow lines from annihilation directly to photons. The latter would represent a very pristine signature of DM, since such spectral features are not expected from any other astrophysical processes. Searches for narrow lines in the Fermi-LAT data were not successful so far -- only upper limits were derived, see e.g. \cite{lines}. This justifies a motivation to develop new gamma-ray telescopes with better spectroscopic capabilities. One such project GAMMA-400 is being developed in Russia now \cite{g400}. Its DM detection capabilities are described in details in \cite{g400-DM}. The estimates of GAMMA-400 sensitivity to the narrow spectral lines around GC presented in \cite{g400-DM} tend slightly towards an optimistic side. Our most recent simulations of GAMMA-400 energy resolution showed that realistically we can expect this sensitivity to be comparable to that of Fermi-LAT in the energy range $\sim$(0.1-100) GeV. At the same time DAMPE has comparable sensitivity too \cite{DAMPE}. Hence the combined (stacked) data analysis from all three telescopes will potentially allow to significantly extend the range of probed diphoton annihilation cross sections.

The situation with the searches of continuum signals from DM annihilation or decay is less ignorant: already a while ago a significant excess of gamma rays around GC was firmly identified (e.g. \cite{abaz12}). It is difficult to explain such diffuse excess by phenomenological  models of emission from CRs, described, e.g., in the monographs \citep{gin64,ber90} and implemented in sophisticated numerical codes such as GALPROP \citep{mos98,jon18}.

The excess is spherically symmetric, visible up to $\sim 10\degree$ from GC and peaks around 2 GeV. It can be well fitted by annihilating WIMPs with the mass of several decades of GeV, the annihilation cross section around the thermal value and steep density profile (e.g. \cite{mauro21,hoop11}). However alternative explanations through various astrophysical mechanisms indeed exist. Thus the excess can be caused by the unusual population of $\sim$ 1000 millisecond pulsars in the Galactic bulge - e.g. \cite{macias19}. The pulsar interpretation is probably considered to be the most plausible now. Other astrophysical interpretations include the emission from molecular clouds, the base of Fermi bubbles, CR burst in the past etc. Also in our opinion the secondary emissions from millisecond pulsars, produced by inverse Compton and bremsstrahlung of $e^{\pm}$ ejected by pulsars, also deserve an attention. Thus the DM interpretation is challenged by both -- its own systematics (e.g. \citet{cal15}) and existence of many competing opportunities.

Then almost all the mentioned opportunities were challenged at some point by the non-trivial observational fact: the abnormal excessive emission was identified in other regions of the Galactic plane \cite{acker17}. This may indicate that the excess source is concentrated around the Galactic plane rather than around GC. This in turn supports the molecular cloud (MC) origin of the excess originally proposed in \cite{de16}, since the clouds indeed reside along all the plane. This creates a motivation to investigate such interesting opportunity further, which comprises the subject of our work. The next section describes in details our model of CR interaction with MCs, which may modify the MC emission spectra through peculiar way leading to imitation of some excessive emission at GeV energies.

\section{Modeling the CR interaction with molecular clouds}

The authors \citet{de16} found that the GC excess is correlated with the distribution of MCs. They tested two hypothesis of the GeV-excess: the real excess hypothesis assuming the excess is caused by DM annihilation, and the seeming excess hypothesis assuming that the “GeV-excess” is related to a hypothetical depletion of gamma rays from MCs below $\approx$ 2 GeV, as it is directly observed in the Central Molecular Zone (CMZ). The authors \cite{de16} did not model in details the origin of the emission depletion below 2 GeV, however stated that it is most likely caused by a magnetic cutoff of cosmic rays in MCs. We have developed the latter idea further.

Generally speaking (see e.g. \cite{acker17,de16}) the hadronic gamma-ray emission is presented by the two components: $\pi^0$ gamma-ray emission from the diffuse gas around the GC and $\pi^0$ gamma-ray emission from MCs. The investigation by \cite{de16} showed that the derived spatial distribution of the emission from MCs correlated nicely with the molecular hydrogen (traced by CO) distribution. Moreover, they found that the gamma-ray emission in the direction on the GC is fully determined by the MC component, where a huge mass of $5 \cdot 10^7 M_\odot$ of molecular hydrogen is concentrated, while other components have a minor effect there. \citet{de16} speculated that this 2 GeV spectral bump is formed due to depletion of the density of CRs with certain energies inside the clouds.

\citet{ivlev} and \citet{dog3} abandoned the phenomenological model of CR propagation and conducted more realistic non-linear simulations of gamma-ray emission from MCs, which produces naturally the "GeV excess" through the "MHD-barrier", which reflects CRs with energies $\lesssim 10$ GeV and does not allow them to penetrate into the clouds. We assumed that the apparent GeV excess does not have a real source, but rather can be explained through an unaccounted modification of MC spectra at certain energies and overall renormalization of the spectra of all emission components. MC spectra gets depleted below few GeV due to a self-excited MHD turbulence by CRs, when the density of CRs inside the cloud is depleted by magnetic fluctuations around the MCs. A flux of CRs propagating through the diffuse ionized gas can excite self-generated magnetic fluctuations (MHD waves). As for the required renormalization of the emission components, it is not large taking into account that the overall intensity of the GC excess does not exceed $\approx$ 10\% of the total gamma-ray intensity within 10$\degree$ from GC (see e.g. \cite{acker17}). Such renormalization by $\approx$ 10\% could be made relatively easily accounting the systematic uncertainties related to the gas emissivity. Let us discuss now in details our model of the modification of MC spectra.

The following system of nonlinear kinetic equations describes processes of CR penetration into MCs \citep[see details of the equations in][]{ivlev} in one-dimensional approximation. The spectrum of CRs $f(z,E,t)$  and the spectrum of magnetic fluctuations $W$ are described by
\begin{eqnarray}
\frac{\partial f}{\partial t}&&= \frac{\partial }{\partial z} \left( D\frac{\partial f}{\partial z} - v_A f\right) +
\frac{\partial }{\partial p} \left(\frac{dp}{dt}f\right)\label{11}\\
\frac{\partial W}{\partial t}&&= \frac{\partial}{\partial k}\left(\frac{W}{T_{\rm nl}}\right) - \frac{\partial}{\partial z}\left( v_a W\right) + \Gamma W -\nu W \label{22}\,.
\end{eqnarray}
where $p$ is the particle momentum, the spatial diffusion coefficient
\begin{equation}
D\simeq \frac{vB^2}{6\pi^2k^2W}\,,
\end{equation}
 $\Gamma$ is the frequency of CR collisions with the background gas, $v_A$ is the Alfvenic velocity, the rate of stream instability excited by CR is
 \begin{equation}
 \Gamma(k,z)\simeq \pi^2{e^2v_A\over m_pc^2\Omega}pvD{\partial f\over\partial z} \,,
 \end{equation}
$(\partial/\partial k)\left(W/T_{\rm nl}(k)\right)$ describes the nonlinear cascade of magnetic fluctuations and $\nu$ describes waves damping due to ion-neutral friction.

The  flux of CRs $S(E) = v_A f - D\frac{\partial f}{\partial z}$ propagating through the envelope and entering the dense interior of the cloud is derived as (see Eqs.(\ref{11})-(\ref{22}))
 \begin{equation}
S(E) = \frac{v_Af_{IS}(E)}{1-\delta e^{-\eta_0(E)}}
\label{Sflux}
\end{equation}
where
\begin{equation}
\eta_0(E)=v_A\int\limits_0^H{dz\over D(E,z)} \,,
\end{equation}
$H$ is the size of the envelope and $f_{IS}$ is the CR spectrum in the the interstellar medium. Coordinate $z=0$ corresponds to the boundary between ISM and the envelope, while $z=H$ corresponds to the conditional boundary between the envelope and the dense interior. The parameter $\delta < 1$ in the Eq. (\ref{Sflux}) describes opacity of the cloud and depends on its column density $N_H$ (see \cite{dog3} for details).

The expression for $\eta_0$ can be estimated from a balance of the growth rate between MHD wave excitation and damping rates due to ion-neutral collisions,  $\Gamma = \nu$,
which leads to the following equation for the $\eta_0$:
\begin{equation}
1-\delta e^{-\eta_0(E)} \propto \delta vpf_{IS}(E) \,.
\label{eq:ex_damp_balance}
\end{equation}

Since $\delta < 1$ and $f_{IS}(E)$ decreases with energy quite steeply, there is a certain value of $E = E_{ex}$, where $\eta_0(E_{ex}) = 0$. And at higher energies the balance can no longer be maintained. This threshold energy, $E_{ex}$, is a boundary between range of free penetration into the cloud core and the range of self modulated CRs. At sufficiently high energies, $E>E_{ex}$, the CR flux is not affected by MHD-turbulence and converges to the regime of the free-streaming (since $\eta_0 = 0$ and $D \rightarrow \infty$). 

Within the range of wave excitation by CRs,  $E < E_{ex}$, the spectrum of MHD fluctuations is provided by excitation-damping balance, described by Eq. (\ref{eq:ex_damp_balance}). As result below $\sim E_{ex}$ a universal spectrum of CRs $f_p(E) \propto E^{-1}$ is formed inside the dense regions of MCs. This spectrum is independent of external parameters outside the cloud. Above $\sim E_{ex}$ the spectrum in the core is the same as in the intercloud medium:
\begin{equation}
f_p(E)\propto
 \left\{
\begin{array}{ll}
E^{-1}\,, & {\rm if}~E \lesssim E_{ex}\,, \\
E^{-\gamma_p}\,, & {\rm if}~E \gtrsim E_{ex}\,,
\end{array}\right.
\label{Np}
\end{equation}
where $\gamma_p$ is the spectral index of the CRs spectrum in the intercloud medium of the CMZ.

Self-excited MHD turbulence may also impact the gamma-ray emission produced by relativistic electrons, since relativistic protons and electrons interact with turbulence in the same way. This is also very relevant for GeV excess interpretations, since the bremsstrahlung and inverse Compton emission from relativistic electrons may comprise significant part of the total emission intensity in the GC region (see e.g. \cite{yus13,acker17}). Hence any minor modifications of those emission components may have a big impact on the GC excess amplitude and spectrum.

Even if the intensity of bremsstrahlung from electrons is lower as compared to the emission from proton-proton collisions, the energy break in the spectrum of bremsstrahlung emission is expected to be located at higher energies. Therefore visible effect of CRs self-modulation on bremsstrahlung emission should appear at much lower column densities.

\begin{figure}[H]
\includegraphics[width=13.5 cm]{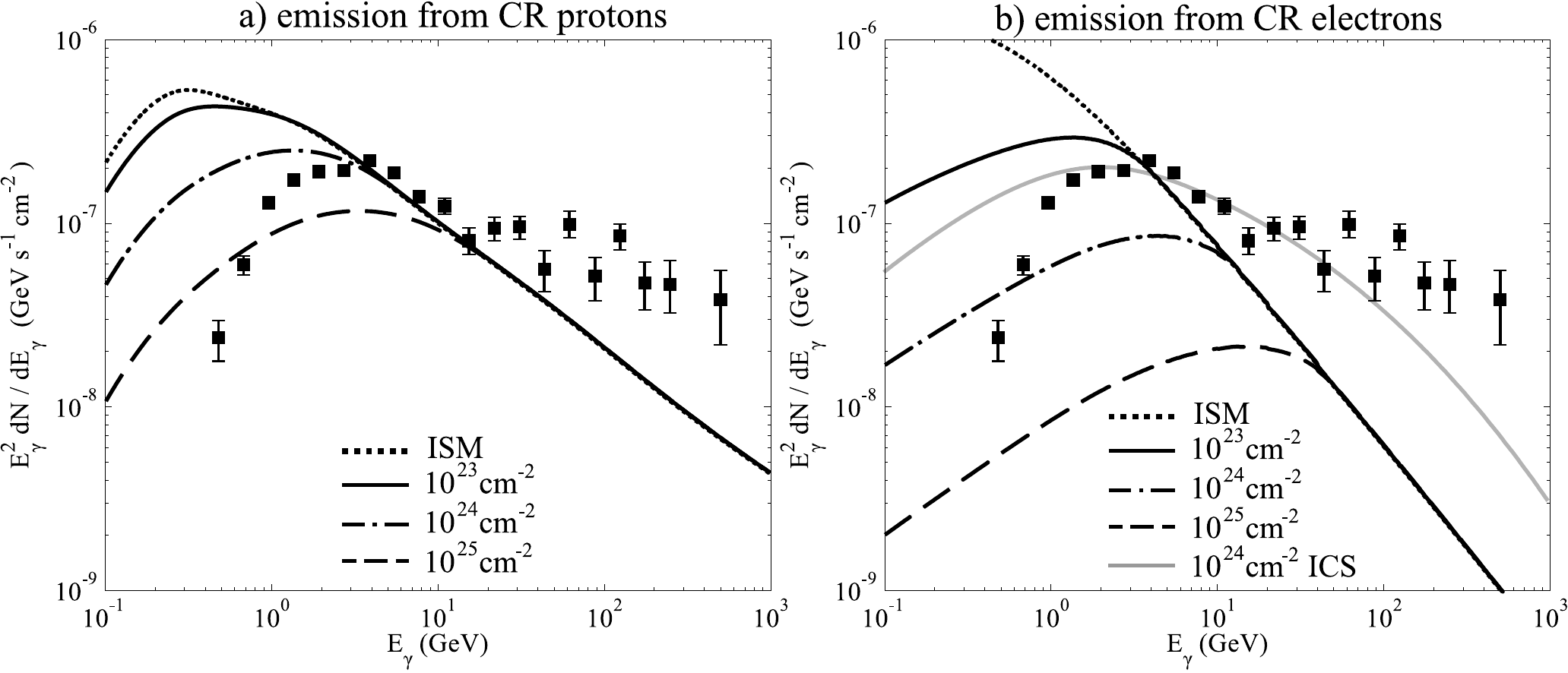}
\caption{Gamma-ray emission from particles inside the clouds for different values of the cloud column density. (\textbf{a}) Emission from relativistic protons (\textbf{b}) Bremsstrahlung emission from relativistic electrons. ICS emission due to scattering of optical photons is also displayed for comparison. The data points represents the GC excess spectrum according to \citet{acker17}. \label{fig1}}
\end{figure}

To illustrate the potential influence of self-generated MHD turbulence on the observed spectra of gamma-ray emission we showed in Figure \ref{fig1} spectra of proton-proton emission $\left(\frac{dN}{dE}\right)_{pp}$ and electron bremsstrahlung emission $\left(\frac{dN}{dE}\right)_{br}$ from the particles inside the cloud. For comparison we displayed the experimental data points of the GeV excess taken from \citet{acker17}. 

We used the following expressions to calculate the gamma-ray emission spectra:
\begin{equation}
\label{eq:gamma_general}
\left(\frac{dN}{dE_\gamma}\right)_i = \frac{M_i c}{4\pi R_{GC}^2 m_p}\int dE~f_j(E) \left(\frac{d\sigma(E,E_\gamma)}{dE_\gamma}\right)_{i}\,,
\end{equation}
where $i$ stands for either $pp$ or $br$ and $j$ is either $p$ in case of proton-proton interactions or $e$ in the case of electron bremsstrahlung. $M$ is the total mass of of clouds emitting gamma-rays, $R_{GC} = 8$ kpc is the distance to the Galactic center and $m_p$ is the mass of the proton. The differential cross section of the photon production in proton-proton interactions is taken from \citet{ppcross_1}, the bremsstrahlung cross section is from \citet{blum70}. 

Eq. (\ref{eq:gamma_general}) assumes that CR spectrum does not change significantly throughout the cloud. This assumption is based on the fact that clouds absorb only a small portion of the penetrating relativistic protons and electrons, therefore their density is almost constant in the dense regions of the clouds. CR spectra experience significant spatial variations in the diffuse envelope, where modulation process takes place. However mass of that envelope is negligible as compared to the total mass of the cloud and gamma-ray emission from this envelope can be ignored.

We did not implemented any spatial modeling of the gamma rays in the current work. The main reason is that to produce a correct spatial distribution of the gamma-ray emission we need to know how clouds with different masses and different column densities are distributed in the Galaxy. Indeed, for each cloud there should be a specific value of $E_{ex}$ and therefore to get a correct spectral shape we need to integrate over the distribution of all values of $E_{ex}$. We intend to do this in the future, however here we only want to show possible effects of the CR modulation on the gamma-ray spectrum.

We assumed the CR density in the GC is the same as local CR density \citep{acero16} and took the analytical forms of CR spectra from \citet{ivlev15}. We considered 4 representative values of the cloud column density: 0 (marked in Figure 1 as ISM), $10^{23}$, $10^{24}$ and $10^{25}$ cm$^{-2}$; which according to \citet{dog3} correspond to the following values of $E_{ex}$: 0, 3.4, 12, 43 GeV. 
These column densities can be observed in the following clouds. $N_{H_2} = 10^{23}$ cm$^{-2}$ can correspond to the average column density of CMZ \citep{ferrier07} or to the envelope of Sgr B2 \citep{Huett95}. $N_{H_2} = 10^{24}$ cm$^{-2}$ and $N_{H_2} = 10^{25}$ cm$^{-2}$ correspond to moderate density region and core of Sgr B2 accordingly \citep{Huett95}. One should notice however that Sgr B2 is one of the most massive molecular clouds in the Galaxy, therefore column density as high as $10^{24} - 10^{25}$ cm$^{-2}$ should be considered as a rare occurrence.

The parameters of the ISM were taken as follows: magnetic field strength is $B = 10~\mu$G and the diffuse gas density is $n = 10$ cm$^{-3}$ \citep{ferrier07,ferrier09}. We note that exact value of the magnetic field strength is uncertain especially in the central regions of the Galaxy, however its value does not affect the value of $E_{ex}$ significantly. Similarly to \citet{ivlev} we assumed that the most abundant ion in the envelope is $C^+$ with density of $3\times 10^{-3}$ cm$^{-3}$. The corresponding value of the Alfven velocity in the envelope is $v_{\rm A} = 10^7$ cm/s.

To reproduce the curves shown in Fig. \ref{fig1} we need the following values of the total mass. For proton-proton collisions it is equal to $M_{pp} = 2\times 10^{8} M_\odot$ and for electron bremsstrahlung it equals to $M_{br} = 2\times 10^{9} M_\odot$. These values exceed the total mass of CMZ and even the mass of the Galactic circumnuclear disk (CND) \citep{ferrier07}. The reason for this discrepancy may be due to influence of the local clouds along the line of sight. Indeed we assumed that all molecular clouds are located at the distance of $R = 8$ kpc and ignored the contribution of clouds which are located in the vicinity. In the case of electrons we can also argue that their density in the Galactic center is much higher than their local density \citep{yus13}.

We also plotted the ICS emission from the modulated spectrum of electrons due to scattering of the soft photons with energy of 1 eV. To calculate it we used the Eq. (\ref{eq:gamma_general}), where we replaced factor $\frac{M}{m_p}$ with $\varepsilon V$. Here $\varepsilon$ is average energy density of soft photons in the Galaxy and $V$ is the total volume involved in the emission. The differential cross section of the ICS emission is taken from \citet{blum70}.

If we assume that $\varepsilon = 10$ eV$\cdot$cm$^{-3}$ \citep{porter17}, we need the volume to be equal to $V = 7\times 10^{65}$ cm$^{3}$ to match the data points, and assuming that region is spherical that corresponds to its radius of about $r=2$ kpc. This is larger than the size of the the CMZ and CND, however we can argue that soft photon density is much higher in the central regions of the Galaxy.

As one can see from Figure \ref{fig1}, modulation of CR flux due to self-generated turbulence indeed produces the deficit of gamma-ray photons at the required energies below several GeV. Thus our model is able to justify the explanation of GC excess \citep[see][]{de16}. 

One can assume that the analysis performed by e.g. \citet{acker17} used standard assumptions (incorporated by GALPROP) to calculate the emissivity of MCs and did not take into account self-modulation of the CRs. Therefore they overestimated the intensity of all the key emission components coming from MCs: hadronic due to proton-proton interactions, bremsstrahlung and ICS from electrons - at the energies below several GeV. 

Although the intercloud diffuse gas emissivity in our model is not modified, the modification of the part of total emission coming from MCs is potentially able to change the ratio of all components significantly. The deficit of gamma-ray intensity from MCs below several GeV would require in turn an increase of the GC excess intensity at those energies in order to preserve the whole balance. The GC excess brightening at low energies in turn would make its spectrum more monotonous, i.e. less bumpy. And then this kind of power-law spectrum is relatively easier to ``absorb'' into the overall diffuse gas or point sources (both resolved and unresolved) spectra, which have the similar shape close to a power law, by their slight renormalization towards higher levels. Such procedure would remove the GC excess as a separate emission component merging the former with the known components.

Finally we would like to notice that our model curves in Figure \ref{fig1} are not intended to fit the GC excess directly. They just illustrate that the reasonable parameter values for MCs and ISM are able to produce the spectral break in intensity of MCs at the energy near the GC excess peak allowing to ``brighten'' the excess below this energy up to the state (ideally), when the whole excess spectrum is close to a monotonous power law.

\section{Conclusions and discussion}

We elaborated the mechanism of the modification of emission spectra of MCs, which according to \cite{de16} can fit the mysterious GeV excess around GC. Our mechanism represents a viable alternative to other possible excess explanations like annihilating DM, millisecond pulsars etc. In order to investigate the nature of the GC excess further and finally establish its exact source the astrophysical community needs in new high-quality gamma-ray data. Thus we crucially need to increase angular and energy resolutions of our instruments around the energies $E_\gamma \sim 1$ GeV with respect to capabilities of Fermi-LAT. This will allow to conduct the resolved imaging of the CMZ region \cite{dog3} and discriminate better the spectra of various emission components there. At the same time we still can not exclude the DM origin of the excess. Hence further investigations may eventually lead to DM discovery. That is why we have big expectations from the currently planned new gamma-ray telescopes: {\it GAMMA-400} \citep[][]{g400-DM}, {\it e-ASTROGAM} \citep[][]{ang21}, {\it AMEGO} \citep[][]{McEnery19} and others.

\vspace{6pt}
%\authorcontributions{Conceptualization, D.O.C., V.A.D. and A.V.I; investigation, D.O.C and A.E.E; writing---original draft preparation, D.O.C, V.A.D and A.E.E; writing---review and editing, D.O.C., A.E.E. and A.V.I.; supervision, V.A.D. All authors have read and agreed to the published version of the manuscript.}%mdpi: For research articles with several authors, a short paragraph specifying their individual contributions must be provided. The following statements should be used ``Conceptualization, X.X. and Y.Y.; methodology, X.X.; software, X.X.; validation, X.X., Y.Y. and Z.Z.; formal analysis, X.X.; investigation, X.X.; resources, X.X.; data curation, X.X.; writing---original draft preparation, X.X.; writing---review and editing, X.X.; visualization, X.X.; supervision, X.X.; project administration, X.X.; funding acquisition, Y.Y. All authors have read and agreed to the published version of the manuscript.'', please turn to the  \href{http://img.mdpi.org/data/contributor-role-instruction.pdf}{CRediT taxonomy} for the term explanation. Authorship must be limited to those who have contributed substantially to the work~reported.

\funding{This research was supported by Russian Science Foundation via the Project 20-12-00047.}

\end{paracol}
%%%%%%%%%%%%%%%%%%%%%%%%%%%%%%%%%%%%%%%%%%
\reftitle{References}


%\externalbibliography{yes}
%\bibliography{your_external_BibTeX_file}

\begin{thebibliography}{999}

\bibitem[{Adriani} {et~al.}(2009){Adriani}, {Barbarino}, {Bazilevskaya},
  {Bellotti}, {Boezio}, {Bogomolov}, {Bonechi}, {Bongi}, {Bonvicini}, {Bottai},
  {Bruno}, {Cafagna}, {Campana}, {Carlson}, {Casolino}, {Castellini}, {de
  Pascale}, {de Rosa}, {de Simone}, {di Felice}, {Galper}, {Grishantseva},
  {Hofverberg}, {Koldashov}, {Krutkov}, {Kvashnin}, {Leonov}, {Malvezzi},
  {Marcelli}, {Menn}, {Mikhailov}, {Mocchiutti}, {Orsi}, {Osteria}, {Papini},
  {Pearce}, {Picozza}, {Ricci}, {Ricciarini}, {Simon}, {Sparvoli},
  {Spillantini}, {Stozhkov}, {Vacchi}, {Vannuccini}, {Vasilyev}, {Voronov},
  {Yurkin}, {Zampa}, {Zampa}, and {Zverev}]{adr17}
{Adriani}, O.; {Barbarino}, G.C.; {Bazilevskaya}, G.A.; {Bellotti}, R.;
  {Boezio}, M.; {Bogomolov}, E.A.; {Bonechi}, L.; {Bongi}, M.; {Bonvicini}, V.;
  {Bottai}, S.; et al.
\newblock {An anomalous positron abundance in cosmic rays with energies
  1.5--100 GeV}.
\newblock {\em \nat} {\bf 2009}, {\em 458},~607--609,
%  \href{http://xxx.lanl.gov/abs/0810.4995}{{\normalfont
%  [arXiv:astro-ph/0810.4995]}}.
\newblock
  doi:{\changeurlcolor{black}\href{https://doi.org/10.1038/nature07942}{\detokenize{10.1038/nature07942}}}.

\bibitem[{Bertone} and {Silk}(2010)]{Bertone-book}
{Bertone}, G.; {Silk}, J. {Particle dark matter}.
\newblock In {\em Particle Dark Matter: Observations, Models and Searches};
  {Bertone}, G., Ed.; Publisher: Cambridge University Press, The Edinburgh Building, Cambridge CB2 8RU, UK, 2010; %mdpi: please add publisher and its location
  p.~3.
	%Cambridge University Press 2010
	%Cambridge University Press
	%The Edinburgh Building, Cambridge CB2 8RU, UK

\bibitem[{Ackermann} {et~al.}(2015){Ackermann}, {Ajello}, {Albert},
  {Anderson}, {Atwood}, {Baldini}, {Barbiellini}, {Bastieri}, {Bellazzini},
  {Bissaldi}, {Blandford}, {Bloom}, {Bonino}, {Bottacini}, {Brandt}, {Bregeon},
  {Bruel}, {Buehler}, {Buson}, {Caliandro}, {Cameron}, {Caputo}, {Caragiulo},
  {Caraveo}, {Cecchi}, {Charles}, {Chekhtman}, {Chiang}, {Chiaro}, {Ciprini},
  {Claus}, {Cohen-Tanugi}, {Conrad}, {Cuoco}, {Cutini}, {D'Ammando}, {de
  Angelis}, {de Palma}, {Desiante}, {Digel}, {Di Venere}, {Drell},
  {Drlica-Wagner}, {Favuzzi}, {Fegan}, {Franckowiak}, {Fukazawa}, {Funk},
  {Fusco}, {Gargano}, {Gasparrini}, {Giglietto}, {Giordano}, {Giroletti},
  {Godfrey}, {Gomez-Vargas}, {Grenier}, {Grove}, {Guiriec}, {Gustafsson},
  {Hewitt}, {Hill}, {Horan}, {J{\'o}hannesson}, {Johnson}, {Kuss}, {Larsson},
  {Latronico}, {Li}, {Li}, {Longo}, {Loparco}, {Lovellette}, {Lubrano},
  {Malyshev}, {Mayer}, {Mazziotta}, {McEnery}, {Michelson}, {Mizuno},
  {Moiseev}, {Monzani}, {Morselli}, {Murgia}, {Nuss}, {Ohsugi}, {Orienti},
  {Orlando}, {Ormes}, {Paneque}, {Pesce-Rollins}, {Piron}, {Pivato},
  {Rain{\`o}}, {Rando}, {Razzano}, {Reimer}, {Reposeur}, {Ritz},
  {S{\'a}nchez-Conde}, {Schulz}, {Sgr{\`o}}, {Siskind}, {Spada}, {Spandre},
  {Spinelli}, {Tajima}, {Takahashi}, {Thayer}, {Tibaldo}, {Torres}, {Tosti},
  {Troja}, {Vianello}, {Werner}, {Winer}, {Wood}, {Wood}, {Zaharijas},
  {Zimmer}, and {Fermi LAT Collaboration}]{lines}
{Ackermann}, M.; {Ajello}, M.; {Albert}, A.; {Anderson}, B.; {Atwood}, W.B.;
  {Baldini}, L.; {Barbiellini}, G.; {Bastieri}, D.; {Bellazzini}, R.;
  {Bissaldi}, E.; et al.
\newblock {Updated search for spectral lines from Galactic dark matter
  interactions with pass 8 data from the Fermi Large Area Telescope}.
\newblock {\em \prd} {\bf 2015}, {\em 91},~122002,
%  \href{http://xxx.lanl.gov/abs/1506.00013}{{\normalfont
%  [arXiv:astro-ph.HE/1506.00013]}}.
\newblock
  doi:{\changeurlcolor{black}\href{https://doi.org/10.1103/PhysRevD.91.122002}{\detokenize{10.1103/PhysRevD.91.122002}}}.

\bibitem[{Galper} {et~al.}(2017){Galper}, {Suchkov}, {Topchiev},
  {Arkhangelskaja}, {Arkhangelskiy}, {Bakaldin}, {Gusakov}, {Dalkarov},
  {Egorov}, {Zverev}, {Kadilin}, {Leonov}, {Naumov}, {Runtso}, {Kheymits}, and
  {Yurkin}]{g400}
{Galper}, A.M.; {Suchkov}, S.I.; {Topchiev}, N.P.; {Arkhangelskaja}, I.V.;
  {Arkhangelskiy}, A.I.; {Bakaldin}, A.V.; {Gusakov}, Y.V.; {Dalkarov}, O.D.;
  {Egorov}, A.E.; {Zverev}, V.G.; et al.
\newblock {Precision Measurements of High-Energy Cosmic Gamma-Ray Emission with
  the GAMMA-400 Gamma-Ray Telescope}.
\newblock {\em Phys. Atom. Nuclei} {\bf 2017}, {\em 80},~1141--1145.
\newblock
  doi:{\changeurlcolor{black}\href{https://doi.org/10.1134/S1063778817060096}{\detokenize{10.1134/S1063778817060096}}}.

\bibitem[{Egorov} {et~al.}(2020){Egorov}, {Topchiev}, {Galper}, {Dalkarov},
  {Leonov}, {Suchkov}, and {Yurkin}]{g400-DM}
{Egorov}, A.E.; {Topchiev}, N.P.; {Galper}, A.M.; {Dalkarov}, O.D.; {Leonov},
  A.A.; {Suchkov}, S.I.; {Yurkin}, Y.T.
\newblock {Dark matter searches by the planned gamma-ray telescope GAMMA-400}.
\newblock {\em \jcap} {\bf 2020}, {\em 11},~049,
%  \href{http://xxx.lanl.gov/abs/2005.09032}{{\normalfont
%  [arXiv:astro-ph.HE/2005.09032]}}.
\newblock
  doi:{\changeurlcolor{black}\href{https://doi.org/10.1088/1475-7516/2020/11/049}{\detokenize{10.1088/1475-7516/2020/11/049}}}.

\bibitem[{Abazajian} and {Kaplinghat}(2012)]{abaz12}
{Abazajian}, K.N.; {Kaplinghat}, M.
\newblock {Detection of a gamma-ray source in the Galactic Center consistent
  with extended emission from dark matter annihilation and concentrated
  astrophysical emission}.
\newblock {\em \prd} {\bf 2012}, {\em 86},~083511,
%  \href{http://xxx.lanl.gov/abs/1207.6047}{{\normalfont
%  [arXiv:astro-ph.HE/1207.6047]}}.
\newblock
  doi:{\changeurlcolor{black}\href{https://doi.org/10.1103/PhysRevD.86.083511}{\detokenize{10.1103/PhysRevD.86.083511}}}.

\bibitem[{Ginzburg} and {Syrovatskii}(1964)]{gin64}
{Ginzburg}, V.L.; {Syrovatskii}, S.I.
\newblock {\em {The Origin of Cosmic Rays}}; Macmillan: New York, NY, USA,  1964.

\bibitem[{Berezinskii} {et~al.}(1990){Berezinskii}, {Bulanov}, {Dogiel},
  {Ginzburg}, and {Ptuskin}]{ber90}
{Berezinskii}, V.S.; {Bulanov}, S.V.; {Dogiel}, V.A.; {Ginzburg}, V.L.;
  {Ptuskin}, V.S.
\newblock {\em {Astrophysics of Cosmic Rays}}; North Holland: Amsterdam, The Netherlands,  1990.

\bibitem[{Moskalenko} and {Strong}(1998)]{mos98}
{Moskalenko}, I.V.; {Strong}, A.W.
\newblock {Production and Propagation of Cosmic-Ray Positrons and Electrons}.
\newblock {\em \apj} {\bf 1998}, {\em 493},~694--707,
%  \href{http://xxx.lanl.gov/abs/astro-ph/9710124}{{\normalfont
%  [arXiv:astro-ph/astro-ph/9710124]}}.
\newblock
  doi:{\changeurlcolor{black}\href{https://doi.org/10.1086/305152}{\detokenize{10.1086/305152}}}.

\bibitem[{J{\'o}hannesson} {et~al.}(2018){J{\'o}hannesson}, {Porter}, and
  {Moskalenko}]{jon18}
{J{\'o}hannesson}, G.; {Porter}, T.A.; {Moskalenko}, I.V.
\newblock {The Three-dimensional Spatial Distribution of Interstellar Gas in
  the Milky Way: Implications for Cosmic Rays and High-energy Gamma-ray
  Emissions}.
\newblock {\em \apj} {\bf 2018}, {\em 856},~45,
%  \href{http://xxx.lanl.gov/abs/1802.08646}{{\normalfont
%  [arXiv:astro-ph.HE/1802.08646]}}.
\newblock
  doi:{\changeurlcolor{black}\href{https://doi.org/10.3847/1538-4357/aab26e}{\detokenize{10.3847/1538-4357/aab26e}}}.

\bibitem[{Di Mauro} and {Winkler}(2021)]{mauro21}
{Di Mauro}, M.; {Winkler}, M.W.
\newblock {Multimessenger constraints on the dark matter interpretation of the
  Fermi-LAT Galactic center excess}. \emph{arXiv}
\newblock  {\bf 2021}, arXiv:2101.11027,
%  \href{http://xxx.lanl.gov/abs/2101.11027}{{\normalfont
%  [arXiv:astro-ph.HE/2101.11027]}}.

\bibitem[{Hooper} and {Goodenough}(2011)]{hoop11}
{Hooper}, D.; {Goodenough}, L.
\newblock {Dark matter annihilation in the Galactic Center as seen by the Fermi
  Gamma Ray Space Telescope}.
\newblock {\em Phys. Lett. B} {\bf 2011}, {\em 697},~412--428,
%  \href{http://xxx.lanl.gov/abs/1010.2752}{{\normalfont
%  [arXiv:hep-ph/1010.2752]}}.
\newblock
  doi:{\changeurlcolor{black}\href{https://doi.org/10.1016/j.physletb.2011.02.029}{\detokenize{10.1016/j.physletb.2011.02.029}}}.

\bibitem[{Macias} {et~al.}(2019){Macias}, {Horiuchi}, {Kaplinghat},
  {Gordon}, {Crocker}, and {Nataf}]{macias19}
{Macias}, O.; {Horiuchi}, S.; {Kaplinghat}, M.; {Gordon}, C.; {Crocker}, R.M.;
  {Nataf}, D.M.
\newblock {Strong evidence that the galactic bulge is shining in gamma rays}.
\newblock {\em \jcap} {\bf 2019}, {\em 2019},~042,
%  \href{http://xxx.lanl.gov/abs/1901.03822}{{\normalfont
%  [arXiv:astro-ph.HE/1901.03822]}}.
\newblock
  doi:{\changeurlcolor{black}\href{https://doi.org/10.1088/1475-7516/2019/09/042}{\detokenize{10.1088/1475-7516/2019/09/042}}}.

\bibitem[{Calore} {et~al.}(2015){Calore}, {Cholis}, and {Weniger}]{cal15}
{Calore}, F.; {Cholis}, I.; {Weniger}, C.
\newblock {Background model systematics for the Fermi GeV excess}.
\newblock {\em \jcap} {\bf 2015}, {\em 2015},~038,
%  \href{http://xxx.lanl.gov/abs/1409.0042}{{\normalfont
%  [arXiv:astro-ph.CO/1409.0042]}}.
\newblock
  doi:{\changeurlcolor{black}\href{https://doi.org/10.1088/1475-7516/2015/03/038}{\detokenize{10.1088/1475-7516/2015/03/038}}}.

\bibitem[{Ackermann} {et~al.}(2017){Ackermann}, {Ajello}, {Albert},
  {Atwood}, {Baldini}, {Ballet}, {Barbiellini}, {Bastieri}, {Bellazzini},
  {Bissaldi}, {Blandford}, {Bloom}, {Bonino}, {Bottacini}, {Brandt}, {Bregeon},
  {Bruel}, {Buehler}, {Burnett}, {Cameron}, {Caputo}, {Caragiulo}, {Caraveo},
  {Cavazzuti}, {Cecchi}, {Charles}, {Chekhtman}, {Chiang}, {Chiappo}, {Chiaro},
  {Ciprini}, {Conrad}, {Costanza}, {Cuoco}, {Cutini}, {D'Ammando}, {de Palma},
  {Desiante}, {Digel}, {Di Lalla}, {Di Mauro}, {Di Venere}, {Drell}, {Favuzzi},
  {Fegan}, {Ferrara}, {Focke}, {Franckowiak}, {Fukazawa}, {Funk}, {Fusco},
  {Gargano}, {Gasparrini}, {Giglietto}, {Giordano}, {Giroletti}, {Glanzman},
  {Gomez-Vargas}, {Green}, {Grenier}, {Grove}, {Guillemot}, {Guiriec},
  {Gustafsson}, {Harding}, {Hays}, {Hewitt}, {Horan}, {Jogler}, {Johnson},
  {Kamae}, {Kocevski}, {Kuss}, {La Mura}, {Larsson}, {Latronico}, {Li},
  {Longo}, {Loparco}, {Lovellette}, {Lubrano}, {Magill}, {Maldera}, {Malyshev},
  {Manfreda}, {Martin}, {Mazziotta}, {Michelson}, {Mirabal}, {Mitthumsiri},
  {Mizuno}, {Moiseev}, {Monzani}, {Morselli}, {Negro}, {Nuss}, {Ohsugi},
  {Orienti}, {Orlando}, {Ormes}, {Paneque}, {Perkins}, {Persic},
  {Pesce-Rollins}, {Piron}, {Principe}, {Rain{\`o}}, {Rando}, {Razzano},
  {Razzaque}, {Reimer}, {Reimer}, {S{\'a}nchez-Conde}, {Sgr{\`o}}, {Simone},
  {Siskind}, {Spada}, {Spandre}, {Spinelli}, {Suson}, {Tajima}, {Tanaka},
  {Thayer}, {Tibaldo}, {Torres}, {Troja}, {Uchiyama}, {Vianello}, {Wood},
  {Wood}, {Zaharijas}, {Zimmer}, and {Fermi LAT Collaboration}]{acker17}
{Ackermann}, M.; {Ajello}, M.; {Albert}, A.; {Atwood}, W.B.; {Baldini}, L.;
  {Ballet}, J.; {Barbiellini}, G.; {Bastieri}, D.; {Bellazzini}, R.;
  {Bissaldi}, E.; et~al.
\newblock {The Fermi Galactic Center GeV Excess and Implications for Dark
  Matter}.
\newblock {\em \apj} {\bf 2017}, {\em 840},~43,
%  \href{http://xxx.lanl.gov/abs/1704.03910}{{\normalfont
%  [arXiv:astro-ph.HE/1704.03910]}}.
\newblock
  doi:{\changeurlcolor{black}\href{https://doi.org/10.3847/1538-4357/aa6cab}{\detokenize{10.3847/1538-4357/aa6cab}}}.
	
\bibitem[{de Boer} {et~al.}(2017){de Boer}, {Bosse}, {Gebauer}, {Neumann},
  and {Biermann}]{de16}
{de Boer}, W.; {Bosse}, L.; {Gebauer}, I.; {Neumann}, A.; {Biermann}, P.L.
\newblock {Molecular clouds as origin of the Fermi gamma-ray GeV excess}.
\newblock {\em \prd} {\bf 2017}, {\em 96},~043012,
 % \href{http://xxx.lanl.gov/abs/1707.08653}{{\normalfont
%  [arXiv:astro-ph.HE/1707.08653]}}.
\newblock
  doi:{\changeurlcolor{black}\href{https://doi.org/10.1103/PhysRevD.96.043012}{\detokenize{10.1103/PhysRevD.96.043012}}}.

\bibitem[{Ivlev} {et~al.}(2018){Ivlev}, {Dogiel}, {Chernyshov}, {Caselli},
  {Ko}, and {Cheng}]{ivlev}
{Ivlev}, A.V.; {Dogiel}, V.A.; {Chernyshov}, D.O.; {Caselli}, P.; {Ko}, C.M.;
  {Cheng}, K.S.
\newblock {Penetration of Cosmic Rays into Dense Molecular Clouds: Role of
  Diffuse Envelopes}.
\newblock {\em \apj} {\bf 2018}, {\em 855},~23,
%  \href{http://xxx.lanl.gov/abs/1802.02612}{{\normalfont
%  [arXiv:astro-ph.HE/1802.02612]}}.
\newblock
  doi:{\changeurlcolor{black}\href{https://doi.org/10.3847/1538-4357/aaadb9}{\detokenize{10.3847/1538-4357/aaadb9}}}.

\bibitem[{Dogiel} {et~al.}(2018){Dogiel}, {Chernyshov}, {Ivlev}, {Malyshev},
  {Strong}, and {Cheng}]{dog3}
{Dogiel}, V.A.; {Chernyshov}, D.O.; {Ivlev}, A.V.; {Malyshev}, D.; {Strong},
  A.W.; {Cheng}, K.S.
\newblock {Gamma-Ray Emission from Molecular Clouds Generated by Penetrating
  Cosmic Rays}.
\newblock {\em \apj} {\bf 2018}, {\em 868},~114,
%  \href{http://xxx.lanl.gov/abs/1810.05821}{{\normalfont
%  [arXiv:astro-ph.HE/1810.05821]}}.
\newblock
  doi:{\changeurlcolor{black}\href{https://doi.org/10.3847/1538-4357/aae827}{\detokenize{10.3847/1538-4357/aae827}}}.

\bibitem[{Yusef-Zadeh} {et~al.}(2013){Yusef-Zadeh}, {Hewitt}, {Wardle},
  {Tatischeff}, {Roberts}, {Cotton}, {Uchiyama}, {Nobukawa}, {Tsuru}, {Heinke},
  and {Royster}]{yus13}
{Yusef-Zadeh}, F.; {Hewitt}, J.W.; {Wardle}, M.; {Tatischeff}, V.; {Roberts},
  D.A.; {Cotton}, W.; {Uchiyama}, H.; {Nobukawa}, M.; {Tsuru}, T.G.; {Heinke},
  C.; et al.
\newblock {Interacting Cosmic Rays with Molecular Clouds: A Bremsstrahlung
  Origin of Diffuse High-energy Emission from the Inner
  2{\textdegree} {$\times$} 1{\textdegree} of the Galactic Center}.
\newblock {\em \apj} {\bf 2013}, {\em 762},~33,
%  \href{http://xxx.lanl.gov/abs/1206.6882}{{\normalfont
%  [arXiv:astro-ph.HE/1206.6882]}}.
\newblock
  doi:{\changeurlcolor{black}\href{https://doi.org/10.1088/0004-637X/762/1/33}{\detokenize{10.1088/0004-637X/762/1/33}}}.

\bibitem[{Kafexhiu} {et~al.}(2014){Kafexhiu}, {Aharonian}, {Taylor}, and
  {Vila}]{ppcross_1}
{Kafexhiu}, E.; {Aharonian}, F.; {Taylor}, A.M.; {Vila}, G.S.
\newblock {Parametrization of gamma-ray production cross sections for p p
  interactions in a broad proton energy range from the kinematic threshold to
  PeV energies}.
\newblock {\em \prd} {\bf 2014}, {\em 90},~123014,
%  \href{http://xxx.lanl.gov/abs/1406.7369}{{\normalfont
%  [arXiv:astro-ph.HE/1406.7369]}}.
\newblock
  doi:{\changeurlcolor{black}\href{https://doi.org/10.1103/PhysRevD.90.123014}{\detokenize{10.1103/PhysRevD.90.123014}}}.

\bibitem[{Blumenthal} and {Gould}(1970)]{blum70}
{Blumenthal}, G.R.; {Gould}, R.J.
\newblock {Bremsstrahlung, Synchrotron Radiation, and Compton Scattering of
  High-Energy Electrons Traversing Dilute Gases}.
\newblock {\em Rev. Mod. Phys.} {\bf 1970}, {\em 42},~237--271.
\newblock
  doi:{\changeurlcolor{black}\href{https://doi.org/10.1103/RevModPhys.42.237}{\detokenize{10.1103/RevModPhys.42.237}}}.

\bibitem[{Acero} {et~al.}(2016){Acero}, {Ackermann}, {Ajello}, {Albert},
  {Baldini}, {Ballet}, {Barbiellini}, {Bastieri}, {Bellazzini}, {Bissaldi},
  {Bloom}, {Bonino}, {Bottacini}, {Brandt}, {Bregeon}, {Bruel}, {Buehler},
  {Buson}, {Caliandro}, {Cameron}, {Caragiulo}, {Caraveo}, {Casandjian},
  {Cavazzuti}, {Cecchi}, {Charles}, {Chekhtman}, {Chiang}, {Chiaro}, {Ciprini},
  {Claus}, {Cohen-Tanugi}, {Conrad}, {Cuoco}, {Cutini}, {D'Ammando}, {de
  Angelis}, {de Palma}, {Desiante}, {Digel}, {Di Venere}, {Drell}, {Favuzzi},
  {Fegan}, {Ferrara}, {Focke}, {Franckowiak}, {Funk}, {Fusco}, {Gargano},
  {Gasparrini}, {Giglietto}, {Giordano}, {Giroletti}, {Glanzman}, {Godfrey},
  {Grenier}, {Guiriec}, {Hadasch}, {Harding}, {Hayashi}, {Hays}, {Hewitt},
  {Hill}, {Horan}, {Hou}, {Jogler}, {J{\'o}hannesson}, {Kamae}, {Kuss},
  {Landriu}, {Larsson}, {Latronico}, {Li}, {Li}, {Longo}, {Loparco},
  {Lovellette}, {Lubrano}, {Maldera}, {Malyshev}, {Manfreda}, {Martin},
  {Mayer}, {Mazziotta}, {McEnery}, {Michelson}, {Mirabal}, {Mizuno}, {Monzani},
  {Morselli}, {Nuss}, {Ohsugi}, {Omodei}, {Orienti}, {Orlando}, {Ormes},
  {Paneque}, {Pesce-Rollins}, {Piron}, {Pivato}, {Rain{\`o}}, {Rando},
  {Razzano}, {Razzaque}, {Reimer}, {Reimer}, {Remy}, {Renault},
  {S{\'a}nchez-Conde}, {Schaal}, {Schulz}, {Sgr{\`o}}, {Siskind}, {Spada},
  {Spandre}, {Spinelli}, {Strong}, {Suson}, {Tajima}, {Takahashi}, {Thayer},
  {Thompson}, {Tibaldo}, {Tinivella}, {Torres}, {Tosti}, {Troja}, {Vianello},
  {Werner}, {Wood}, {Wood}, {Zaharijas}, and {Zimmer}]{acero16}
{Acero}, F.; {Ackermann}, M.; {Ajello}, M.; {Albert}, A.; {Baldini}, L.;
  {Ballet}, J.; {Barbiellini}, G.; {Bastieri}, D.; {Bellazzini}, R.;
  {Bissaldi}, E.; et al.
\newblock {Development of the Model of Galactic Interstellar Emission for
  Standard Point-source Analysis of Fermi Large Area Telescope Data}.
\newblock {\em \apjs} {\bf 2016}, {\em 223},~26,
%  \href{http://xxx.lanl.gov/abs/1602.07246}{{\normalfont
%  [arXiv:astro-ph.HE/1602.07246]}}.
\newblock
  doi:{\changeurlcolor{black}\href{https://doi.org/10.3847/0067-0049/223/2/26}{\detokenize{10.3847/0067-0049/223/2/26}}}.

\bibitem[{Ivlev} {et~al.}(2015){Ivlev}, {Padovani}, {Galli}, and
  {Caselli}]{ivlev15}
{Ivlev}, A.V.; {Padovani}, M.; {Galli}, D.; {Caselli}, P.
\newblock {Interstellar Dust Charging in Dense Molecular Clouds: Cosmic Ray
  Effects}.
\newblock {\em \apj} {\bf 2015}, {\em 812},~135,
%  \href{http://xxx.lanl.gov/abs/1507.00692}{{\normalfont
%  [arXiv:astro-ph.GA/1507.00692]}}.
\newblock
  doi:{\changeurlcolor{black}\href{https://doi.org/10.1088/0004-637X/812/2/135}{\detokenize{10.1088/0004-637X/812/2/135}}}.

\bibitem[{Ferri{\`e}re} {et~al.}(2007){Ferri{\`e}re}, {Gillard}, and
  {Jean}]{ferrier07}
{Ferri{\`e}re}, K.; {Gillard}, W.; {Jean}, P.
\newblock {Spatial distribution of interstellar gas in the innermost 3 kpc of
  our galaxy}.
\newblock {\em \aap} {\bf 2007}, {\em 467},~611--627,
%  \href{http://xxx.lanl.gov/abs/astro-ph/0702532}{{\normalfont
%  [arXiv:astro-ph/astro-ph/0702532]}}.
\newblock
  doi:{\changeurlcolor{black}\href{https://doi.org/10.1051/0004-6361:20066992}{\detokenize{10.1051/0004-6361:20066992}}}.

\bibitem[{Huettemeister} {et~al.}(1995){Huettemeister}, {Wilson},
  {Mauersberger}, {Lemme}, {Dahmen}, and {Henkel}]{Huett95}
{Huettemeister}, S.; {Wilson}, T.L.; {Mauersberger}, R.; {Lemme}, C.; {Dahmen},
  G.; {Henkel}, C.
\newblock {A multilevel study of ammonia in star-forming regions. VI. The
  envelope of Sagittarius B2.}
\newblock {\em \aap} {\bf 1995}, {\em 294},~667--676.

\bibitem[{Ferri{\`e}re}(2009)]{ferrier09}
{Ferri{\`e}re}, K.
\newblock {Interstellar magnetic fields in the Galactic center region}.
\newblock {\em \aap} {\bf 2009}, {\em 505},~1183--1198,
%  \href{http://xxx.lanl.gov/abs/0908.2037}{{\normalfont
%  [arXiv:astro-ph.GA/0908.2037]}}.
\newblock
  doi:{\changeurlcolor{black}\href{https://doi.org/10.1051/0004-6361/200912617}{\detokenize{10.1051/0004-6361/200912617}}}.

\bibitem[{Porter} {et~al.}(2017){Porter}, {J{\'o}hannesson}, and
  {Moskalenko}]{porter17}
{Porter}, T.A.; {J{\'o}hannesson}, G.; {Moskalenko}, I.V.
\newblock {High-energy Gamma Rays from the Milky Way: Three-dimensional Spatial
  Models for the Cosmic-Ray and Radiation Field Densities in the Interstellar
  Medium}.
\newblock {\em \apj} {\bf 2017}, {\em 846},~67,
%  \href{http://xxx.lanl.gov/abs/1708.00816}{{\normalfont
%  [arXiv:astro-ph.HE/1708.00816]}}.
\newblock
  doi:{\changeurlcolor{black}\href{https://doi.org/10.3847/1538-4357/aa844d}{\detokenize{10.3847/1538-4357/aa844d}}}.

\bibitem[{De Angelis} {et~al.}(2021){De Angelis}, {Tatischeff}, {Argan},
  {Brandt}, {Bulgarelli}, {Bykov}, {Costantini}, {Curado da Silva}, {Grenier},
  {Hanlon}, {Hartmann}, {Hernanz}, {Kanbach}, {Kuvvetli}, {Laurent},
  {Mazziotta}, {McEnery}, {Morselli}, {Nakazawa}, {Oberlack}, {Pearce}, {Rico},
  {Tavani}, {von Ballmoos}, {Walter}, {Wu}, {Zane}, {Zdziarski}, and
  {Zoglauer}]{ang21}
{De Angelis}, A.; {Tatischeff}, V.; {Argan}, A.; {Brandt}, S.; {Bulgarelli},
  A.; {Bykov}, A.; {Costantini}, E.; {Curado da Silva}, R.; {Grenier}, I.A.;
  {Hanlon}, L.; et al.
\newblock {Gamma-ray Astrophysics in the MeV Range: The ASTROGAM Concept and
  Beyond}.
\newblock {\em arXiv} {\bf 2021},  arXiv:2102.02460.
%  \href{http://xxx.lanl.gov/abs/2102.02460}{{\normalfont
%  [arXiv:astro-ph.IM/2102.02460]}}.

\bibitem[{McEnery} {et~al.}(2019){McEnery}, {van der Horst}, {Dominguez},
  {Moiseev}, {Marcowith}, {Harding}, {Lien}, {Giuliani}, {Inglis}, {Ansoldi},
  {Stamerra}, {Manousakis}, {Strong}, {Bambi}, {Patricelli}, {Baring},
  {Barrio}, {Bastieri}, {Fields}, {Beacom}, {Beckmann}, {Bednarek}, {Rani},
  {Boggs}, {Bolotnikov}, {Cenko}, {Buckley}, {Grefenstette}, {Hui}, {Pittori},
  {Prescod-Weinstein}, {Shrader}, {Gouiffes}, {Kierans}, {Wilson-Hodge},
  {D'Ammando}, {Castro}, {Kocveski}, {Gasparrini}, {Thompson}, {Williams}, {De
  Angelis}, {Bernard}, {Digel}, {Morcuende}, {Charles}, {Bissaldi}, {Hays},
  {Ferrara}, {Bozzo}, {Grove}, {Wulf}, {Bottacini}, {Caroli}, {Kislat},
  {Oikonomou}, {Giordano}, {Longo}, {Fryer}, {Fukazawa}, {Georganopoulos}, {De
  Nolfo}, {Vianello}, {Kanbach}, {Younes}, {Blumer}, {Hartmann}, {Hernanz},
  {Takahashi}, {Li}, {Agudo}, {Moskalenko}, {Stumke}, {Grenier}, {Smith},
  {Rodi}, {Perkins}, {Gelfand}, {Holder}, {Knodlseder}, {Kopp}, {Lenain},
  {{\'A}lvarez}, {Metcalfe}, {Krizmanic}, {Stephen}, {Hewitt}, {Mitchell},
  {Harding}, {Tomsick}, {Racusin}, {Finke}, {Kargaltsev}, {Klimenko},
  {Krawczynski}, {Smith}, {Kubo}, {Di Venere}, {Marcotulli}, {Lommler},
  {Parker}, {Baldini}, {Foffano}, {Zampieri}, {Tibaldo}, {Petropoulou},
  {Ajello}, {Meyer}, {L{\'o}pez}, {McConnell}, {Boettcher}, {Cardillo},
  {Martinez}, {Kerr}, {Mazziotta}, {McEnery}, {Di Mauro}, {Wood}, {Meyer},
  {Briggs}, {De Becker}, {Lovellette}, {Doro}, {Sanchez-Conde}, {Moss},
  {Mizuno}, {Rib{\'o}}, {Nakazawa}, {Neilson}, {Auricchio}, {Omodei},
  {Oberlack}, {Ohno}, {Orlando}, {Otte}, {Coppi}, {Bloser}, {Zhang}, {Laurent},
  {Pohl}, {Prandini}, {Shawhan}, {Caputo}, {Campana}, {Rando}, {Woolf},
  {Johnson}, {Mignani}, {Walter}, {Ojha}, {da Silva}, {Dietrich}, {Funk},
  {Zane}, {Anton}, {Buson}, {Cutini}, {Saz Parkinson}, {Schirato}, {Griffin},
  {Kaufmann}, {Stawarz}, {Ciprini}, {Del Sordo}, {Jones}, {Guiriec}, {Tajima},
  {Cheung}, {The}, {Venters}, {Porter}, {Linden}, {Barres}, {Paliya},
  {Bozhilov}, {Vestrand}, {Tatischeff}, {Chen}, {Wang}, {Tanaka}, {Uhm},
  {Zhang}, {Zimmer}, {Zoglauer}, and {Wadiasingh}]{McEnery19}
{McEnery}, J.; {van der Horst}, A.; {Dominguez}, A.; {Moiseev}, A.;
  {Marcowith}, A.; {Harding}, A.; {Lien}, A.; {Giuliani}, A.; {Inglis}, A.;
  {Ansoldi}, S.; et al.
\newblock {All-sky Medium Energy Gamma-ray Observatory: Exploring the Extreme
  Multimessenger Universe}.
\newblock  \emph{Bull. Am. Astron. Soc.}  \textbf{2019}, \emph{51},  245.
%  ,  \href{http://xxx.lanl.gov/abs/1907.07558}{{\normalfont
%  [arXiv:astro-ph.IM/1907.07558]}}.


\bibitem{DAMPE}
	{Xu}, Z.L.; {Duan}, K.K.; {Jiang}, W.; {Lei}, S.J.; {Li}, X.; {Shen}, Z.Q.; {Ma}, T.; {Su}, M.; {Yuan}, Q.; {Yue}, C.; et al.
\newblock{Optimal gamma-ray selections for monochromatic line searches with DAMPE}.
\newblock \emph{arXiv} \textbf{2021}, arXiv:2107.13208.



\end{thebibliography}
\def\aap{A\&A}
\def \apjs{ApJS}
\def\apj{ApJ}
\def\jcap{JCAP}
\def\nat{Nature} 
\def\prd{PhRvD}

%=====================================
% References, variant B: internal bibliography
%=====================================

%\begin{thebibliography}{999}

%\bibitem[Author1(year)]{ref-journal}
%Author~1, T. The title of the cited article. {\em Journal Abbreviation} {\bf 2008}, {\em 10}, 142--149.

% \bibitem[Author8(year)]{ref-url}
%Title of Site. Available online: URL (accessed on Day Month Year).
%\end{thebibliography}

% If authors have biography, please use the format below
%\section*{Short Biography of Authors}
%\bio
%{\raisebox{-0.35cm}{\includegraphics[width=3.5cm,height=5.3cm,clip,keepaspectratio]{Definitions/author1.pdf}}}
%{\textbf{Firstname Lastname} Biography of first author}
%
%\bio
%{\raisebox{-0.35cm}{\includegraphics[width=3.5cm,height=5.3cm,clip,keepaspectratio]{Definitions/author2.jpg}}}
%{\textbf{Firstname Lastname} Biography of second author}

% The following MDPI journals use author-date citation: Arts, Econometrics, Economies, Genealogy, Humanities, IJFS, JRFM, Laws, Religions, Risks, Social Sciences. For those journals, please follow the formatting guidelines on http://www.mdpi.com/authors/references
% To cite two works by the same author: \citeauthor{ref-journal-1a} (\citeyear{ref-journal-1a}, \citeyear{ref-journal-1b}). This produces: Whittaker (1967, 1975)
% To cite two works by the same author with specific pages: \citeauthor{ref-journal-3a} (\citeyear{ref-journal-3a}, p. 328; \citeyear{ref-journal-3b}, p.475). This produces: Wong (1999, p. 328; 2000, p. 475)

\end{document}